\documentclass[12pt]{article}
\usepackage{epsf}
\begin{document}
\begin{titlepage}
\title{Remark on Koide's $Z_3$-symmetric parametrization of quark masses}
\author{
{Piotr \.Zenczykowski }\footnote{E-mail: piotr.zenczykowski@ifj.edu.pl}\\
{\em Division of Theoretical Physics}\\
{\em the Henryk Niewodnicza\'nski Institute of Nuclear Physics}\\
{\em Polish Academy of Sciences}\\
{\em Radzikowskiego 152,
31-342 Krak\'ow, Poland}\\
}
\maketitle
\begin{abstract}
The charged lepton masses may be parametrized in a $Z_3$-symmetric language
appropriate to the discussions of Koide's formula. The phase parameter
$\delta_L$ appearing in this parametrization is experimentally
indistinguishable from $2/9$. We analyse Koide's parametrization for the up
($U$) and down ($D$) quarks and argue that the data are suggestive of
 the low-energy values $\delta_U=\delta_L/3=2/27$ and 
 $\delta_D=2\delta_L/3=4/27$.

\end{abstract}

\vfill
\end{titlepage}

Although the Standard Model is extremely successful, 
 various questions concerning elementary particles 
 cannot be answered within it. Among them, notwithstanding
the recent discovery at CERN, 
there is still the issue of particle masses. 
This problem seems to be intimately 
related to the appearance of three generations of leptons and quarks.
Since the understanding of these issues may require completely new ideas, phenomenological
identification
of regularities observed in the pattern of particle masses and mixings 
is of crucial importance. It may provide us with analogues of 
Balmer and Rydberg's formulae and should hopefully lead us to a genuinely
 new physics.\\

{\it 1.} One of the most interesting of such regularities was discovered by Koide
\cite{Koide} (for a brief review see \cite{RiveroGsponer}). It reads:
\begin{equation}
\label{KoideFormula}
\frac{m_e+m_{\mu}+m_{\tau}}{(\sqrt{m_e}+\sqrt{m_{\mu}}+\sqrt{m_{\tau}})^2}=
\frac{1+k_L^2}{3},
\end{equation}
with $k_L$ equal exactly $1$.

In fact, if one plugs into the above formula the central values of 
experimental electron and muon masses \cite{PDG}:
\begin{eqnarray}
m_e(exp)&=&0.510998928(11)~{\rm MeV}, \nonumber \\
\label{electronmuonmassesexp}
m_{\mu}(exp)&=&105.65836715(35)~{\rm MeV},
\end{eqnarray}
one finds (with $k_L=1$) that the larger of the solutions of Eq. (\ref{KoideFormula}) is
\begin{equation}
\label{tauKoideprediction}
m_{\tau}=1776.9689~{\rm MeV},
\end{equation}
to be compared with the experimental $\tau$ mass
\begin{equation}
\label{tauonmassexp}
m_{\tau}(exp)=1776.82\pm 0.16~{\rm MeV}.
\end{equation}

Discussions of this success of Koide's formula are often 
formulated in a $Z_3$-symmetric
language by parametrizing the masses of any three given fermions $f_1,f_2,f_3$
 as  \cite{Koide2,Brannen}:
\begin{eqnarray}
\label{KoideParametrization}
\sqrt{m_j}&=&\sqrt{M_f}\,\left(1+\sqrt{2}\,k_f 
\cos {\left(\frac{2 \pi j}{3}+\delta_f\right)}\right), ~~~~~~~~~(j=1,2,3).
\end{eqnarray} 
In general, with appropriately chosen three parameters (here: the overall mass scale $M_f$ and the pattern parameters 
$ k_f$ and $\delta_f$) one can fit any three-particle spectrum, of course. The above 
 choice of parametrization 
is, however, particularly suited to Koide's formula, since the latter is then
independent of parameter $\delta_f$, as specified
in Eq. (\ref{KoideFormula}). 
That the resulting formula works then for $k_L$
equal exactly 1 (or almost $1$ with very high precision) is truly amazing. \\

{\it 2.} The peculiar feature of formula (\ref{KoideFormula}) is that it works 
at a low
energy scale and not at some high mass scale (like $M_Z$ or $10^{16}~ {\rm GeV}$)
\cite{Koide3,XingZhang,LiMa}. For example, taking the values of charged 
lepton masses as appropriate at the
scale of $M_Z$ ($m_e=0.486755106~{\rm MeV}$, $m_{\mu}=102.740394~{\rm MeV}$,
$m_{\tau}=1746.56~{\rm MeV}$) one finds that $k_L(M_Z)=1.00188$, i.e. it 
 deviates
from unity quite significantly. If that value of $k_L$ worked for physical masses it
would remove much of the excitement Koide's formula generates.
 
Attempts have been made to apply Koide's formula to quarks and neutrinos.
The general conclusion is that the formula does not work there (i.e. $k_f \ne 1$). 
Specifically, using the quark mass
values as appropriate at $\mu = 2~{\rm GeV}$, 
for the down  
quarks ($D$)
one obtains the values of $k_D$ around 1.08, while for the up  quarks ($U$) one gets
$k_U$ around $1.25$ \cite{XingZhang,Kartavtsev,RodejohannZhang} (with the
mathematically allowed region $0 \leq k_f \leq \sqrt{2}$).
Furthermore, for neutrinos $\nu$ one estimates directly from
experiment that
$k_{\nu} \leq 0.81 $ \cite{RodejohannZhang}.
If a higher energy scale
 $\mu$ is taken, the agreement deteriorates further (at the $M_Z$ mass scale 
 one gets $k_D=1.12$ and $k_U=1.29$).\\ 
 
 {\it 3.} Given the success of Koide's parametrization (\ref{KoideParametrization}) 
 with $k_f=1$ for
 charged leptons ($f=L$) and its failure for other fundamental fermions
 ($f=U,D,\nu$),
one should perhaps look in a different direction. In fact, parametrization
(\ref{KoideParametrization}) reveals another miracle in the charged
lepton sector. Namely, using the
experimental values of charged lepton masses one can determine the value of the
phase parameter $\delta_L$.

From Eq. (\ref{KoideParametrization}) one finds
\begin{eqnarray}
\frac{1}{\sqrt{2}}(\sqrt{m_2}-\sqrt{m_1})&=&\sqrt{M_f}\, k_f \sqrt{3} \,\sin
\delta_f,\\
\frac{1}{\sqrt{6}}(2 \sqrt{m_3}-\sqrt{m_2}-\sqrt{m_1})&=&\sqrt{M_f}\, k_f\,
\sqrt{3} \cos \delta_f.
\end{eqnarray}
Since $\delta_f$ is free we may 
assume $m_1\le m_2 \le m_3$ without any loss of generality.
 Then, independently of the value of $k_L$, one gets
 \begin{equation}
 \label{tandeltaL}
 \tan \delta_L
=\frac{\sqrt{3}\,(\sqrt{m_{\mu}}-\sqrt{m_e})}
{2\sqrt{m_{\tau}}-\sqrt{m_{\mu}}-\sqrt{m_e}}.
 \end{equation}
From the experimental values 
of Eqs (\ref{electronmuonmassesexp}, \ref{tauonmassexp}) one then calculates:
\begin{equation}
\delta_L=0.2222324,
\end{equation}
which is extremely close to $\delta_L=2/9$ \cite{Brannen2,Rosen}.
After inverting formula (\ref{tandeltaL}) and assuming $\delta_L=2/9$ one can predict
the value of $\tau$ mass, given the experimental masses of electron and muon. 
The result is:
\begin{equation}
\label{tauRosenprediction}
m_{\tau}=1776.9664 ~{\rm MeV}.
\end{equation}
This is as good a prediction of the tauon mass as that given by the original Koide's 
formula (\ref{tauKoideprediction}). The two
predictions of Eqs (\ref{tauKoideprediction},\ref{tauRosenprediction})
are mutually incompatible, but either of them leads to an excellent
prediction for $m_{\tau}$. They could be made consistent with each other
by allowing extremely
 tiny 
departures of either $\delta_L$ from $2/9$ or $k_L$ from 1.
Keeping in mind the violation
of Koide's formula for quarks and neutrinos, one should perhaps  
try the $\delta_L=2/9$ alternative, for example by maintaining
the values of all $\delta_f$ equal to $\delta_L$. 
An attempt in a similar direction was made by Rosen \cite{Rosen}. 
He keeps the values 
of all $\delta_f$ equal
to $2/9$ and the values of all $k_f$ at $1$. \footnote{Hence, 
due to the above mentioned incompatibility, Rosen cannot simultaneously describe 
the electron and muon masses with maximal precision as required by the data.} 
Then, he constructs a $Z_3$-symmetric model 
which modifies (in a calculable way) the value of $M_{q}$   
for each quark $q_j$ separately
(hence $M_q \to M_{q,j}$). However, then the Koide parametrization of Eq.(\ref{KoideParametrization}) 
ceases to be valid.\\
 
{\it 4.} On the other hand, a different route possible may be taken
which keeps parametrization (\ref{KoideParametrization}) intact. 
Namely, one could
refrain for the time being from the discussions of Koide-like formulas 
(with $k_f \ne 1$). Instead, one might try to analyze the issue of $\delta_f$ in more 
detail. After all, the assumption of $\delta_L=2/9$ yields as good 
a prediction for the tauon mass as the assumption of $k_L=1$.

Thus, the idea is to analyse what the experimental values of 
quark masses tell us about the $\delta_f$
parameters for $f=U,D$. For this simple exercise we take the following
typical set of the values of experimental masses at $\mu =2~{\rm GeV}$ (in MeV):
\begin{eqnarray}
m_u&=&  1.7-3.3,\nonumber \\
m_c&=& 1270^{+70}_{-90}, \nonumber \\
m_t&=&  172000 \pm 1600,\nonumber \\
m_d&=&  4.1-5.8, \nonumber \\
m_s&=&  101^{+29}_{-21},\nonumber \\
\label{expquarkmasses}
m_b&=&  4190^{+180}_{-60}.
\end{eqnarray}
For the discussion of $\delta_f$ we introduce $z_k=\sqrt{m_k/m_3}$
(assuming $m_1 < m_2 < m_3$).
Thus
\begin{equation}
\delta_f = \arctan \left(\sqrt{3}\,\frac{z_2-z_1}{2-z_2-z_1}\right).
\end{equation}
Fig. 1 shows a contour plot of $\delta_f\,(z_1,z_2)$
and the corresponding approximate positions of $(z_1,z_2)$
(together with their errors) as calculated from Eq. (\ref{expquarkmasses})
for the up quarks (marked $U$) and the down quarks (marked $D$).
For comparison with the lepton case the position of $\delta_L$ is also shown 
(marked with a dot $L$). 
Slanted solid lines represent 
constant $\delta ~(=0,\, 1/27,\, 2/27,...,\, 2/9,...)$.
It can be seen that the observed value of $\delta_U$ is consistent 
with $\delta_U= \delta_L/3=2/27$. Thanks to a huge top quark mass, the errors 
are quite small here.
For the down sector the mass hierarchy is not as strong as in the up sector,
and therefore the corresponding errors are much larger. 

\begin{figure}[t]
\caption{Contour plot of $\delta_f\,(z_1,z_2)$ and the relevant points
corresponding to the charged lepton ($L$), up ($U$), and down ($D$) 
quark sectors, as explained in the text.  
}
\label{fig1}
\begin{center}
\epsfxsize=15.0 cm
\mbox{\epsfbox{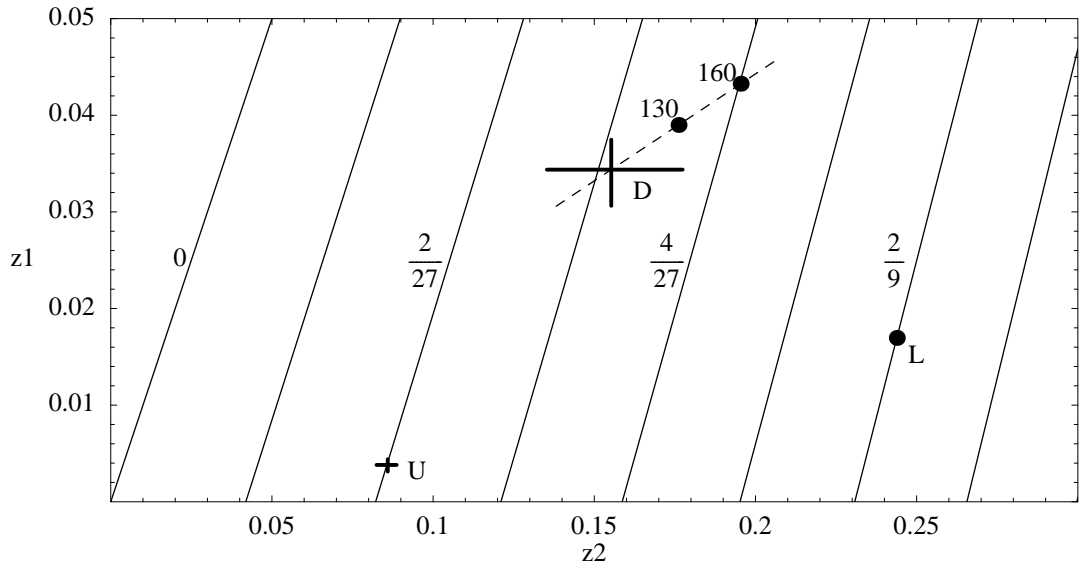}}
\end{center}
 \end{figure}

We have to remember, however, that at $\mu = M_Z$ both $k_L$ 
and $\delta_L$
deviate from their `perfect' values of 1 and 2/9, 
by about 0.2 \% and 0.5 \% respectively 
(the deviation from $2/9$ is virtually indiscernible in the scale
used for the presentation of Fig. 1).
Apparently, we should be interested in the values of current quark masses
at the low energy scale and not at $\mu = 2~{\rm GeV}$ as in Eq.
 (\ref{expquarkmasses}).
For the up quark sector, thanks to the huge mass of the top quark, 
such a change cannot significantly modify 
the position of the corresponding $U$ point in Fig. 1.
This is not the case for the down quarks. In order to show what happens there,
we assumed that $\kappa=m_s/m_d$ is fixed at the value of $\kappa=20.4$ as obtained
 at $\mu = 2~{\rm GeV}$ from  Eq. (\ref{expquarkmasses}) and as also valid at low
 energies when extracted from $\pi$ and $K$ masses (see e.g. \cite{Weinberg}).
The corresponding relation between $z_1$ and $z_2$ is marked in Fig. 1 with a
dashed line.
Two points along this line, corresponding to $m_s=130$ and $160~{\rm MeV}$, are
shown there as well. We observe that 
at the expected low-energy-scale value of the strange quark mass 
(i.e. for $m_s$ around $160~{\rm MeV}$)  
the value of
$\delta_D$ appears to be close to $2 \delta_L/3 =4/27$. The obtained
low-energy-scale values
of $\delta_U$, $\delta_D$ and $\delta_L$ are therefore suggestive of
a nice (even if only fairly approximate) symmetry between the lepton and 
quark sectors, with
the values of $I_3=-1/2$  particle phases  $\delta_L, \delta_D$ depending on
weak hypercharge $Y$ and
given by
\begin{equation}
\delta(I_3=-1/2,Y)=\frac{1}{9}(1+|Y|),
\end{equation}  
and the $I_3=+1/2$  particle phases $\delta_{\nu}, \delta_U$ expected to be
given by
\begin{equation}
\delta(I_3=+1/2,Y)=\frac{1}{9}(1-|Y|),
\end{equation}
together forming an equally-spaced set and satisfying a lepton-quark sum rule $\delta_L+\delta_{\nu}=\delta_U+\delta_D$.

Obviously, due to the possible Majorana mass term, the observed 
masses of neutrinos do not have to realize the pattern $m_1=m_2 < m_3$
suggested by $\delta(I_3=+1/2,|Y|=1)=0$.

\vfill

\vfill

\end{document}